# MindTheDApp: A Toolchain for Complex Network-Driven Structural Analysis of Ethereum-based Decentralised Applications

Giacomo Ibba*, Sabrina Aufiero†, Silvia Bartolucci†, Rumyana Neykova‡,
Marco Ortu*, Roberto Tonelli*, Giuseppe Destefanis‡
*University of Cagliari, Italy
Email: {giacomo.ibba, marco.ortu, roberto.tonelli}@unica.it
†UCL, University College London, UK
Email: {sabrina.aufiero.22, s.bartolucci}@ucl.ac.uk
‡Brunel University London, UK
Email: {rumyana.neykova, giuseppe.destefanis}@brunel.ac.uk
Corresponding author: Giuseppe Destefanis
Email: giuseppe.destefanis@brunel.ac.uk

*Abstract*—This paper presents MindTheDApp, a toolchain designed specifically for the structural analysis of Ethereum-based Decentralized Applications (DApps), with a distinct focus on a complex network-driven approach. Unlike existing tools, our toolchain combines the power of ANTLR4 and Abstract Syntax Tree (AST) traversal techniques to transform the architecture and interactions within smart contracts into a specialized bipartite graph. This enables advanced network analytics to highlight operational efficiencies within the DApp's architecture.

The bipartite graph generated by the proposed tool comprises two sets of nodes: one representing smart contracts, interfaces, and libraries, and the other including functions, events, and modifiers. Edges in the graph connect functions to smart contracts they interact with, offering a granular view of interdependencies and execution flow within the DApp. This network-centric approach allows researchers and practitioners to apply complex network theory in understanding the robustness, adaptability, and intricacies of decentralized systems.

Our work contributes to the enhancement of security in smart contracts by allowing the visualisation of the network, and it provides a deep understanding of the architecture and operational logic within DApps. Given the growing importance of smart contracts in the blockchain ecosystem and the emerging application of complex network theory in technology, our toolchain offers a timely contribution to both academic research and practical applications in the field of blockchain technology.

## I. INTRODUCTION

Solidity is a high-level, statically-typed programming language specifically designed for writing smart contracts on the Ethereum blockchain platform. It incorporates elements of existing languages such as JavaScript and Python, but is tailored to the requirements of blockchain development. One of its standout features is its contract-oriented design, which allows for clear and reusable code structures. This enables developers to create decentralised applications, complex financial mechanisms, and even other blockchains. Its popularity and widespread adoption make Solidity a central subject for study, especially as smart contracts become increasingly integral to blockchain ecosystems. The necessity to analyse Solidity smart contracts is given by two critical aspects: security and structural understanding of Decentralised Applications (DApps). Security vulnerabilities in smart contracts can be dangerous [7], given the immutable nature of blockchain. Being able to parse and analyse the contracts opens the field for identifying such vulnerabilities, allowing for timely remediation. In addition to security, the study of Solidity contracts provides a window into the architecture and operational logic of DApps. These contracts contain the rules and functions that dictate the behaviour of a DApp, making their analysis crucial for understanding how these decentralised systems function. Therefore, an efficient tool for parsing Solidity smart contracts serves a dual purpose: enhancing security and enriching the understanding of DApps. Complex networks theory [8] offer a powerful lens through which to study and understand the behaviour of systems. Complex networks, which could be social, biological, or technological, are characterised by non-trivial topological features that govern the interactions among their individual components [18]. By studying the network structure, researchers can gain valuable insights into emergent system behaviours, such as robustness, adaptability, and efficiency. In the context of blockchain systems, understanding the network interactions within and among smart contracts could provide new perspectives on system vulnerabilities and operational efficiencies [29]. However, to the best of our knowledge, there is limited research on the applicability of complex network theory to the analysis of smart contracts.

S.B., G.D., R.N. and M.O. acknowledge support from the Ethereum foundation grant FY23-1048

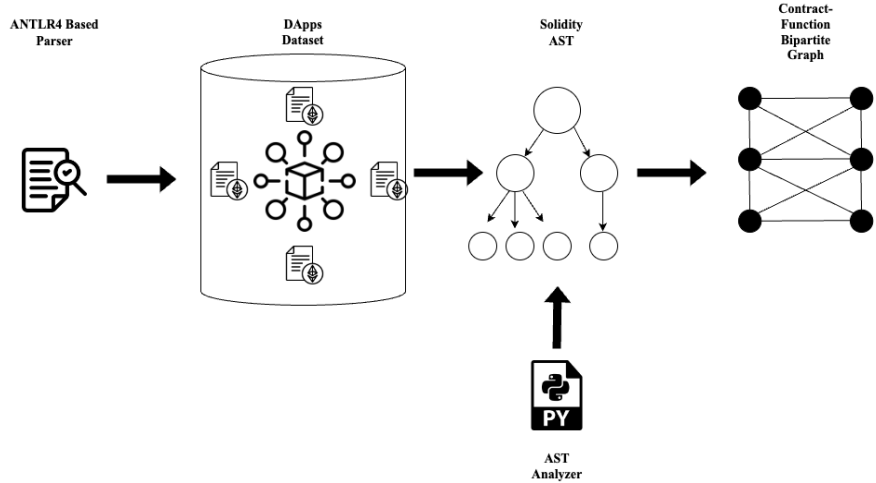

Fig. 1: Toolchain resuming the bipartite graph's creation process

This paper introduces MindTheDApp, a toolchain[1] uniquely designed for the structural analysis of Ethereum-based DApps, emphasizing a complex network-driven approach. Unlike traditional tools that perform structural analysis, our tool uses complex networks analysis techniques to offer an understanding of smart contract interactions. The tool uses ANTLR4 [20] to traverse the Abstract Syntax Tree (AST) of Solidity contracts. This information is then transformed into a specialised bipartite graph, allowing for advanced network analytics that can highlight potential bottlenecks or vulnerable points within the DApp's architecture. In this graph, one set of nodes represents smart contracts, while the other set represents functions. Edges connect functions to the contracts they interact with, providing a comprehensive view of dependencies and flows within the DApp. Our approach goes beyond a simple bipartite structure by offering a detailed, context-specific visualization, making it easier to understand how various contracts and functions are interconnected [2]. The graph produced by the tool is particularly suited for complex network analysis, enabling researchers to study aspects such as contract interdependencies, potential security vulnerabilities, and execution flow within decentralised applications.

In addition to introducing MindTheDApp, this paper also presents an initial dataset of decentralised applications. This dataset spans various categories, including finance, art, gaming, and technology, and serves as a resource for researchers and practitioners. It offers a detailed look into the complex networks of smart contract interactions within DApps, providing a foundation for future studies on DApp structures, network topologies, and potential security concerns.

## II. TOOLCHAIN OVERVIEW

Figure 1 illustrates the workflow of our proposed tool for constructing a complex network-driven bipartite graph representation of a decentralised application. The process begins by tokenising the DApp's smart contracts using the Lexer module. This is followed by generating the Abstract Syntax Tree (AST) representation of the source code through the Parser module. Once the AST is constructed, the Analyzer module scans and extracts key elements relevant to our complex network analysis. These key elements include contracts, functions, interfaces, events, modifiers, and libraries.

In the resulting bipartite graph, nodes represent two distinct categories: the first category comprises functions, events, and modifiers, while the second category includes smart contracts, interfaces, and libraries. An edge exists between a node from the first category and a node from the second category if and only if the function, event, or modifier from the first category calls the corresponding contract, interface, or library in the second category. This includes External calls, which are interactions that involve source code imported from external sources. This structure enables a complex network-driven view of how different components within a DApp interact.

In order to deliver a better understanding of the logic and tool's functionalities, we give an example of a Solidity contract in Figure 2 (a). The code shows a Solidity contract named `Contract3`, which imports and uses `Contract1` and `Contract2`. It implements a function `func3` that calls functions `func1` and `func2` from these imported contracts. In this setup, `Contract3` is the source, and `Contract1` and `Contract2` are the targets. The function `func3` triggers the calls. Both `Contract1` and `Contract2` have similar code structures and make calls to each other and `Contract3`.

The dependency graph generated by the tool is shown in Figure 2 (b). In this specific case, we have six nodes (the three functions and the three contracts), and six edges highlighting the contract calls from the specific functions.

[1]Github Link to the tool

```
1  // SPDX-License-Identifier: MIT
2  pragma solidity ^0.7.6;
3  import "./Contract1.sol";
4  import "./Contract2.sol";
5  contract Contract3 {
6      Contract1 contract1;
7      Contract2 contract2;
8
9      constructor (Contract1 _contract1,
10     Contract2 _contract2) {
11         contract1 = _contract1;
12         contract2 = _contract2;
13     }
14
15     function func3() public {
16         contract1.func1();
17         contract2.func2();
18     }
19 }
```

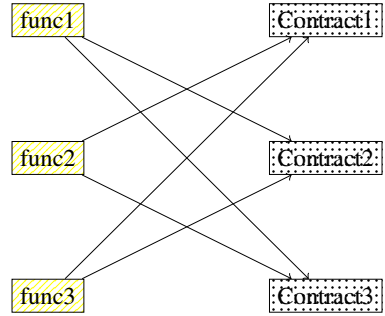

(a)　　　　　　　　　　　　　　(b)

Fig. 2: Example of (a) a solidity contract and (b) the produced bipartite graph

## III. METHODOLOGY AND TOOL ARCHITECTURE

Our tool employs ANTLR4 to perform several key tasks in the smart contract analysis process. After receiving the Solidity source code, ANTLR4's lexer first tokenizes the input, breaking it down into identifiable lexical units. These tokens are then fed to ANTLR4's parser, which organises them into a hierarchical structure, resulting in an Abstract Syntax Tree (AST). We traverse this AST using auto-generated tree walkers, extracting relevant syntactical and semantical information for further analysis. This process is key for producing a bipartite graph that reflects contract and function dependencies in the analysed smart contracts.

Internally, our tool follows a modular architecture comprising several components (highlighted in Figure 3), each responsible for specific tasks. Upon receiving a Solidity contract for analysis, the Lexer and Parser modules tokenize the code. The resulting AST is then passed to the Analysis module, where features like function calls, contract dependencies, and control flows are extracted. This data is represented as a bipartite graph in the Graph module, which is then available for complex network analysis.

The advantages of using ANTLR4 can be summarised as threefold:

- Versatility and Performance: ANTLR4 can handle a wide array of grammar types and employs the efficient LL(*) parsing algorithm, enabling fast and accurate parsing of various languages and formats.
- Error Handling and Management: ANTLR4 excels in identifying and recovering from syntax errors, providing robust error reporting and continuation capabilities.
- Extensibility and Integration: ANTLR4 supports grammar inheritance and can generate parser code for multiple programming languages. It also integrates with build systems like Maven and Gradle, making it a highly adaptable tool for diverse development environments.

Given our focus on building a bipartite graph that depicts function and contract dependencies, the following Solidity constructs are of particular interest to us:

- **Function Definitions and Calls**: These provide insight into the interdependencies between different functions within and across contracts.
- **Modifiers**: Used to change the behaviour of functions, understanding modifiers helps in analysing the conditions and requirements under which functions operate.
- **Events**: These are crucial for tracking changes and interactions, offering a dynamic view of contract activities.
- **Inheritance and Interfaces**: These features help in understanding contract hierarchies, which is essential for analysing dependencies.

## IV. PARSING RULES FOR NETWORK EXTRACTION

The goal of the tool is to create a bipartite graph that illustrates the interactions between functions and contracts, central to Decentralised Applications. The objective is to detect and showcase a function (within the source contract) making a call to another contract (termed the target contract). For this purpose, defining rules to identify valid calls becomes essential. It is important to mention that the source code for the Listener, Lexer, and Parser depends on the grammar used during their creation.

The initial step is to define the kind of nodes in the bipartite graph. In our study, nodes represent functions, contracts, interfaces, and libraries. As a result, any call found in the source code forms a relation *(source,target)*.

For the generation of nodes, specific rules and criteria have been set:

- **Constructor Calls:** The constructor initiates a new instance of a contract. This means that a constructor is seen as a call to the contract itself. For example, if there's a contract named "Bank", the constructor would produce a self-referencing node, the relation (Bank, Bank), with the association being defined by the 'constructor' keyword.
- **Global Scope Calls:** If a contract call is made outside of specific structures like functions, modifiers, or events, and exists within the broad scope of the contract, the link formed from the Source contract to the Target contract is labelled as 'Global'.
- **Self-Reference Calls:** The keyword 'this' denotes the contract itself. Every occurrence of 'this' within a contract is treated as a legitimate call to the contract. For instance, within a "Bank" contract, any 'this' reference would result in a relation (Bank, Bank). The specific function where 'this' appears defines the connection.
- **Cast Operations:** Casting operations to contracts, libraries, or interfaces are treated as valid calls. If there's a "Bank" contract and within it, a cast operation to "ERC20" is executed, the outcome would be a relation (Bank, ERC20). The connecting link is determined by the function where the cast occurs.
- **Calls to Constructs:** Calls made from one contract to libraries, interfaces, or other contract structures (like functions, events, or modifiers) are recognised as valid calls. For example, a "Bank" contract invoking a modifier from another "Vault" contract produces a valid call. This interaction would generate a relation (Bank, Vault), with the specific function in the "Bank" contract that made the call determining the link.
- **External Source Calls:** Calls to contracts, interfaces, or libraries sourced from external platforms, such as GitHub, are labelled as 'External'. So, if a "Bank" contract interacts with an "ERC20" contract imported from GitHub, the relationship would be represented as (Bank, External).

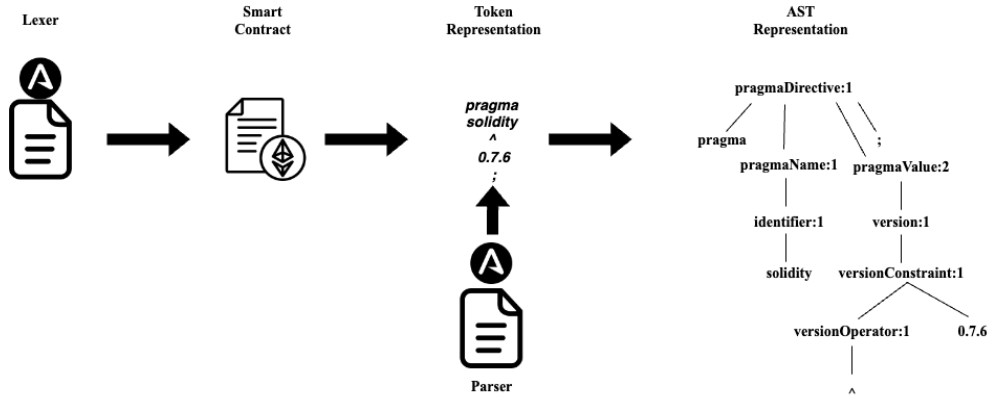

Fig. 3: Toolchain of the ANTLR architecture

Following these rules ensures that the nodes are created accurately and consistently, representing the interactions within decentralised applications effectively.

### A. Parser Definition and Top-Down AST Exploration

This section provides a brief overview of our approach to traversing the Abstract Syntax Tree (AST) of a Solidity smart contract, as shown in Figure 4. We employ a specialised listener classes generated from the Solidity grammar [2]. ANTLR provides support for two tree-walking mechanisms in its runtime library – parse-tree listeners and visitors. We employ the former since (1) it offers a more efficient tree traversal and (2) it is applicable to traversals that do not alter the parse trees, as in our case. In a nutshell, a parse-tree listener interface responds to events triggered by the built-in tree walker. The methods in a listener class are callbacks. The listeners receive notification of events like `startTreeNode` and `endTreeNode`.

Key elements in the `contractDefinition` branch include functions, state variables, events, and modifiers, among others. The listener classes facilitate the traversal of these elements, extracting relevant information that contributes to our complex network-driven bipartite graph analysis.

For isntance, traversing the `ifStatement` branch of the AST involves understanding the boolean conditions that guide the code flow. Similarly, loops like `forStatement` and conditional constructs like `tryStatement` offer unique challenges. These constructs may include further branches that require recursive exploration, as they could involve additional function calls or contract interactions.

### B. Assembly Code Analysis

In Solidity, assembly code allows developers to interact more closely with the Ethereum Virtual Machine (EVM). While Solidity provides a high-level interface for contract creation, assembly code can be used for certain optimizations or EVM-specific behaviors.

Analyzing assembly code within the Abstract Syntax Tree (AST) is different from Solidity code. Figure **??** illustrates the `inlineAssemblyStatement` branch in the AST, encapsulating all statements in the Solidity assembly construct. These statements can include identifiers, blocks, expressions, local definitions, assignments, among others.

[2]https://github.com/solidityj/solidity-antlr4/blob/master/Solidity.g4

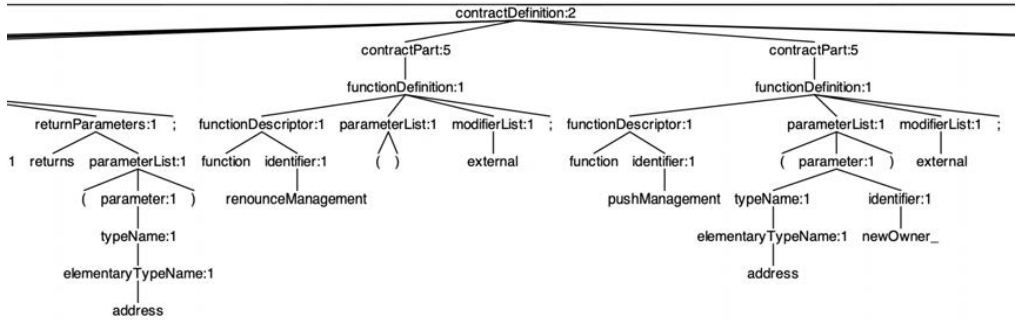

Fig. 4: Abstract Syntax Tree example of a smart contract

Assembly code has a simpler structure than Solidity, making its analysis more straightforward. For instance, all assembly statements can be examined using just `assemblyExpression`, `assemblyLocalDefinition`, and `assemblyAssignment`, which cover variable definitions, assignments, and expressions like function calls.

## V. CONTRACT CALLS (GRAPH NODES) EXTRACTION

This section explains how to extract contract calls from a smart contract's source code. Our primary focus is on identifying all constructors and instances of the **this** keyword to capture self-referential contract calls. Detecting calls to other contracts dispersed throughout the code poses a challenge. These calls can appear within functions, modifiers, events, and custom errors. However, we opt to exclude calls within embedded assembly code due to their rare occurrence and minimal impact on our study.

In our examination of Events and Modifiers, we aim to find contract calls, uses of functions from other contracts, contract objects, references to **this**, and typecasting operations targeting contracts, interfaces, or libraries. Contract calls can also occur within boolean conditions of control structures like **if**, **for**, **while**, and **do-while** statements.

We conduct a systematic examination of all software constructs, covering various types of expressions like variable declarations and assignments. Sometimes, certain constructs might not be present in the DApp's source code, usually because they are imported from external sources like GitHub. In such cases, we extract the source code directly from the relevant web page, generate a unique Abstract Syntax Tree for that contract, and analyze it. Due to the added computational requirements and web page dependency, we label these interactions as "External calls."

*1) Example of Contract Calls within a Function:* In the example below, the function `renounceManagement()` calls a modifier `onlyPolicy()` and emits an event `OwnershipPushed`.

```
1  function renounceManagement() public virtual
       override onlyPolicy() {
2      emit OwnershipPushed(_owner, address(0));
3      _owner = address(0);
4  }
```

Both the modifier and the event come from the same contract, `Ownable`. Invoking the modifier and triggering the event add specific function calls to the sequence, forming pairs involving the contract `Ownable`.

In this example, both the event and the modifier are defined within the same contract, `Ownable`. The function `renounceManagement` is defined within the `Ownable` contract as well, and in this case, we have two different contract calls in which the source and target contract overlap. The first call provides `Ownable` as the source and target contract, `renounceManagement` as the source function, and the modifier `onlyPolicy` being part of the call chain. The second call provides `Ownable` as the source and target contract as well, the `renounceManagement` as the source function and the event `OwnershipPushed` incorporated into the sequence.

In another example, a function called `markdown` includes several contract calls. The function starts by calling `IUniswapV2Pair(_pair).getReserves()` and stores the result in a tuple. The key part is the cast operation, leading to a sequence of function calls involving the `getReserves()` method. Following this, the function has an `ifStatement` containing the expression `IUniswapV2Pair(_pair).token0() == SGT`. This expression also includes a contract call and should be broken down for further analysis.

```
1  function markdown( address _pair )
2  external view returns ( uint ) {
3      (uint reserve, uint reserve1, ) = IUniswapV2Pair
           ( _pair ).getReserves();
4      uint reserve;
5      if ( IUniswapV2Pair( _pair ).token0() == SGT ) {
6          reserve = reserve1;
7      } else {
8          reserve = reserve;
9      }
10     return reserve.mul( 2 * ( 10 ** IERC20( SGT ).
           decimals() ) ).div( getTotalValue( _pair ) )
           ;
11 }
```

The function concludes with a return statement that includes another contract call, casting to `IERC20`. This adds another function call sequence involving `decimals()`, `div()`, and `getTotalValue()`.

To summarize, this function includes three primary contract calls:

- `IUniswapV2Pair(_pair).getReserves()`
- `IUniswapV2Pair(_pair).token0()`
- `IERC20(SGT).decimals()`

The source function, in this specific example, is the function `markdown`, which is the function where the three calls are encapsulated. The source contract is the one that defines the function markdown, while `IUniswapV2Pair` and `IERC20` are the two target contracts. The following functions `getReserves`, `token0`, and `decimals` build the call chain for the three different contract calls.

Lastly, the function `_mint()` in the `ERC20` contract triggers the `Transfer` event from the `IERC20` interface.

```solidity
function _mint(address account_, uint256 amount_)
    internal virtual {
    require(account_ != address(0), "ERC20: mint to
        the zero address");
    _beforeTokenTransfer(address(this), account_,
        amount_);
    _totalSupply = _totalSupply.add(amount_);
    _balances[account_] = _balances[account_].add(
        amount_);
    emit Transfer(address(this), account_, amount_);
}
```

In this case, the function `_mint` is the source function, the contract `ERC20` (that defines the _mint function) is the source contract, and the contract `IERC20`, which defines the `Transfer` event is the target contract. Moreover, the event takes as input the `address(this)` parameter, which refers to the contract itself (in this specific case `ERC20`). The tool considers the `this` as a valid contract call, and consequentially the `ERC20` contract is both source and target contract, the `_mint` function as the source function, and the `Transfer` event is incorporated into the sequence.

This section has walked through multiple examples to illustrate how contract calls within functions are identified and analysed based on our extraction rules. These examples cover different scenarios, including function calls, modifiers, and events, to give a comprehensive view of how contract interactions occur.

## VI. Dataset and Evaluation

We have collected 3093 smart contracts from 26 DApps belonging to different domains and we have generated their respective dependency graphs. The dataset generated by MindTheDApp serves as one of the contributions of this paper. It includes Decentralized Applications (DApps) from various categories, as recommended by Ethereum.org. The dataset provides an initial source for researchers and developers interested in analyzing DApp structures, identifying patterns, and studying the network topology of these applications.

- **Financial**: These DApps focus on crypto-based financial services such as lending, borrowing, and interest accumulation.
- **Art and Collectibles**: This category emphasizes digital ownership and revenue for artists, providing investment opportunities for enthusiasts.
- **Gaming**: These applications offer interactive entertainment, featuring virtual worlds and valuable in-game collectibles.
- **Gambling**: In this category, users can engage in various betting activities, ranging from classic casino games to blockchain-specific prediction markets.
- **Technology**: These DApps aim to decentralize developer tools and integrate crypto-economic systems into existing technologies.

Table 1 provides a summary of the Decentralized Applications included in the dataset.

The columns in the table are defined as follows:

- **Name**: The name of the decentralized application (DApp).
- **Category**: The specific category to which the DApp belongs.
- **Macro Category**: The broader category under which the DApp falls.
- **Number of Smart Contracts**: The total number of smart contracts that form the basis of the DApp.
- **Contract Calls**: The count of contract calls identified by the analyzer.
- **Everyday Users**: The percentage of total users who engage with the particular DApp on a daily basis.
- **Outflow/Inflow**: This column shows the percentage of currency that is both deposited into and withdrawn from the exchange, respectively.

### A. Evaluation

This section reports on some of our preliminary findings and illustrates the performance and accuracy of the tool.

*1) Performance:* To test the applicability of MindTheDApp, we ran it on 728 applications constituted by 25077 smart contracts from the DAppScan dataset [3], which is a curated repository built to assess the performances of smart contracts vulnerability detection tools. We successfully extracted the dependencies within the DAppScan dataset, and here we report the performance of the tool on the executed applications.

We conducted our experiments on a MacBook Air with an Apple M1 processor with 8 cores, 8 GB of RAM, and 256 GB of SSD with macOs Monterey 12.6.6. All the Solidity compiler versions are locally installed in case of a needed version switch, and python (3.8.13), and npm (9.6.7) have been used to run code and install packages respectively.

The results revealed the efficacy of the tool in properly scanning and generating a bipartite graph for the sample of decentralised applications. Trivially, the execution time is strongly conditioned by the DApp's dimension. The biggest DApp of the dataset considering the number of smart contracts (596 SCs), required 65.16686 seconds for the scanning and graph generation process, while the smallest (8 SCs) required 0.55405 seconds. The average execution time is 12.60385 seconds.

---

[3]https://github.com/InPlusLab/DAppSCAN

| Name | Category | Macro Category | # of Contracts | Contract Calls | Everyday Users | Outflow/Inflow |
|---|---|---|---|---|---|---|
| Async | Art and Fashion | Art and Collectibles | 7 | 46 | 9.91% | 8.15%/0.321% |
| Foundation | Art and Fashion | Art and Collectibles | 59 | 218 | 37.09% | 5.65%/0.0599% |
| Super Rare | Art and Fashion | Art and Collectibles | 20 | 64 | 10.62% | 3.3%/0.216% |
| Marble Cards | Digital Collectibles | Art and Collectibles | 38 | 229 | 13.79% | 10.34%/X |
| Rarible | Digital Collectibles | Art and Collectibles | 252 | 722 | 27.16% | 10.86%/1.01% |
| Seaport | Digital Collectibles | Art and Collectibles | 243 | 2477 | 27.34% | 3.4%/0.152% |
| Audius | Music | Art and Collectibles | 39 | 136 | X | X |
| Etherisc | Insurance | Financial | 94 | 486 | X | X |
| NexusMutual | Insurance | Financial | 197 | 606 | X | X |
| Balancer | Investments | Financial | 409 | 1037 | 10.59% | 7.55%/0.188% |
| PoolTogether | Investments | Financial | 55 | 184 | 17.04% | 5.56%/0.515% |
| SetToken | Investments | Financial | 293 | 2020 | | |
| Aave | Lending and Borrowing | Financial | 241 | 628 | 8.39% | 5.48%/1.12% |
| Compound | Lending and Borrowing | Financial | 37 | 292 | 9.03% | 6.05%/1.4% |
| PWN | Lending and Borrowing | Financial | 48 | 113 | 10.48% | 11.29%/X |
| TornadoCash | Payments | Financial | 11 | 40 | X | X |
| 1Inch | Token Swaps | Financial | 8 | 12 | X | X |
| Uniswap | Token Swaps | Financial | 145 | 273 | 23.89% | 10.41%/0.144% |
| Loopring | Trading and Prediction Markets | Financial | 596 | 3022 | 15.46% | 12.55%/2.31% |
| Polymarket | Trading and Prediction Markets | Financial | 83 | 211 | X | X |
| Axie Infinity | Competition | Gaming | 25 | 128 | 5.25% | 4.82%/2.06% |
| Dark Forest | Competition | Gaming | 28 | 313 | X | X |
| Gods Unchained | Competition | Gaming | 25 | 128 | 21.1% | 10.19%/2.18% |
| Crypto Voxels | Virtual World | Gaming | 8 | 211 | 9.15% | 4.21%/0.2% |
| Ethereum Name Service | Utilities | Technology | 116 | 286 | 28.7% | 8.3%/0.152% |

TABLE I: Summary of Decentralized Applications

*2) Example Analyses:* To demonstrate the applicability of MindTheDapp, Figure 5 presents an example of a filtered function network of Ethersic, one of the DApps in our dataset, presented in Table I.

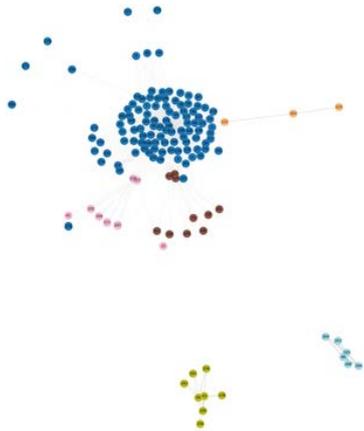

Fig. 5: Filtered network of functions for the Etherisc DApp

Figure 5 showcases how the tool can reveal the complex network of contracts and their interactions within a DApp.

In our study of 26 DApps (Table I) from various categories, we found notable patterns in function and contract interactions. Most functions call between 1 to 4 contracts, suggesting task distribution among multiple functions. All analyzed dApps exhibit high modularity, with modularity coefficients ranging from 0.21 to 0.92, indicating the presence of distinct, non-interconnected components. This pattern is consistent across dApps of different sizes and categories. Metrics like diameter, average path length, and clustering coefficient also display consistency across dApps, hinting at common development patterns.

We employ a disparity filter to isolate the most crucial interactions within the function and contract networks. A disparity filter is a network simplification technique that retains only statistically significant edges, thereby revealing the 'backbone' of a complex network. This approach allows us to focus on the most impactful relationships between functions and contracts, providing a clearer, more meaningful representation of the network's core structure. The use of this filter helps us to distill complex network data into a more manageable form, making it easier to identify key patterns and vulnerabilities.

After applying a disparity filter, we observed that function networks retain about 55% of their nodes, while contract networks shrink dramatically to about 12% of their original size. We define a 'filtered function network' as a projection from the original bipartite graph, where each node represents a function and edges are formed based on certain projection rules capturing interactions between functions. Similarly, a 'filtered contract network' is another projection from the same bipartite graph, but in this case, each node represents a contract, and edges are formed based on interactions between contracts. Both types of networks aim to highlight the specific interplay of functions or contracts within decentralized applications.

Lastly, our resilience analysis shows that targeted removal of high-betweenness nodes can quickly fragment the largest connected component, unlike random removal. This reveals the network's vulnerability to specific disruptions.

*B. Potential Usages*

MindTheDApp offers several avenues for further analysis and study. For example, the tool could be used for:
- Identifying key contracts that serve as hubs in the network, which could be critical points for security evaluation.
- Studying the flow of contract calls to identify potential bottlenecks or inefficiencies in a DApp.
- Comparing the network structures of DApps across different categories to identify common patterns or unique features.

## VII. TOOL APPLICATION AND OUTPUT ANALYSIS

Our tool is able to analyse and extract contract calls from a selected sample of decentralised applications. After scanning the DApp, the tool produces a CSV file named after the application, which contains five columns:
- **File**: Specifies the name of the smart contract.
- **Source Contract**: Identifies the contract where the function calls the target contract.
- **Source Function**: Notes the function that calls the target contract.
- **Target Contract**: Lists the contract called by the source contract.
- **Chain**: If the target contract is called after a chain of function calls, then the whole chain of calls is reported.

Our analysis shows that financial decentralised applications generally have a higher number of contract calls and are typically larger in terms of the number of smart contracts composing the application. MindTheDapp effectively extracts key elements like modifiers and event calls.

We chose to omit external dependencies to concentrate on analyzing the intrinsic structure of a DApp in isolation. This approach allows us to provide a more focused and meaningful representation of the application's network topology. By doing so, we aim to understand the internal interactions, dependencies, and potential bottlenecks within a specific DApp, which are often more relevant for developers and researchers interested in optimizing or securing that particular application.

Including external dependencies would widen the scope of our analysis, potentially diluting the insights gained about the DApp itself. For example, if external dependencies such as common contracts like ERC20 were included in the analysis, they would likely emerge as central nodes in the network graph. While these nodes may be important in the broader Ethereum ecosystem, their centrality could distract from the unique characteristics and vulnerabilities of the DApp being studied. Therefore, our tool, MindTheDapp, aims to offer a more precise, application-specific view of the DApp's internal network structure.

## VIII. THREATS TO VALIDITY

In this section, we outline potential threats to the validity of our research, addressing issues that could affect both the generalizability and applicability of our findings.

**Lack of Cross-Platform Comparison:** Our study is confined to DApps within the Ethereum ecosystem. This narrow focus hampers our understanding of decentralised applications more broadly, as a comparative analysis across different platforms could reveal key similarities or differences within the same DApp categories. Our tool, however, is not platform-specific and can analyse any Solidty contract independently of the platform on which it is deployed.

**Temporal Scope Limitations:** While our research includes popular Ethereum DApps, it lacks a temporal dimension. A more comprehensive study would incorporate DApps developed at various stages of Ethereum's lifecycle, from its early years to the present. Such a comparison could yield valuable insights into the evolving structure and categorization of DApps over time.

**Missing Dependencies:** In our analysis, external dependencies like GitHub imports are labeled as 'External,' obscuring the original contract names. External contracts could offer specific patterns that highlight similarities or differences between DApps.

**Parser Testing Scope:** Our tool underwent testing on a dataset of 728 applications. While this sample size allowed us to identify and address some tool limitations, more extensive testing on a larger dataset is required to further validate the tool's efficiency and effectiveness. For example, initial testing did not account for empty contract declarations, leading to the extraction of None-type objects, an issue that has since been resolved.

## IX. RELATED WORK

Research and development in the areas of smart contracts have seen a surge in recent years. In this section, we discuss previous work that has laid the groundwork for our study and highlight the gaps that our research aims to fill.

*A. Smart Contract Analysis*

Numerous studies have targeted smart contract analysis, mainly to highlight security vulnerabilities [5], [14], [16], [23], [24], [26], [28]. These works have employed a range of methods, from static and dynamic analysis to formal verification and machine learning-based techniques [14], [23], [28]. However, such methods often focus on isolated types of vulnerabilities and lack a broad understanding of the smart contract's functionality [16], [23]. For instance, some studies aim to identify specific kinds of attacks like consensus protocol attacks, smart contract code bugs, operating system malware, or fraudulent users [24]. Others direct their attention to particular analytical aspects such as gas consumption or opcode analysis [15], [19]. While these efforts contribute valuable insights into specific vulnerabilities or aspects, they fall short of offering a thorough perspective on smart contract functionality and vulnerabilities.

To fill this gap, it is important to develop more comprehensive analytical tools. One promising avenue is the application of complex networks to the study of DApps. Complex network analysis can offer additional perspectives on the structure of smart contract interactions, highlighting potential bottlenecks or vulnerable points that may not be evident through conventional analytical methods. Hence, there is a need for integrated approaches that can offer a holistic view of smart contracts and their interactions within decentralised applications.

### B. DApps and Solidity

The Ethereum platform and its native language, Solidity, have been the subject of numerous studies. These studies have explored various aspects of decentralised applications (DApps) and smart contracts on the Ethereum blockchain. Additionally, they have examined how the community of developers influences the perception of the platform [3]. For example, Wu et al. [27] conducted a comprehensive empirical study of 995 Ethereum DApps, analyzing transaction logs to gain insights into the structure and behaviors of DApps. They highlighted the rapid development and wide adoption of DApps in various domains.

Bhargavan et al. [4] focused on the formal verification of Solidity contracts using the F programming language, aiming to prevent bugs and vulnerabilities. In addition to empirical studies and formal verification, researchers have also explored tools and techniques for analyzing Solidity contracts. Hajdu et al. [12] proposed a novel approach for analyzing Solidity contracts, evaluating its semantics and comparing it to other analysis tools. Gao et al. [11] developed a tool called SmartEmbed, which effectively identifies instances of repetitive Solidity code in smart contracts. Furthermore, the adoption and implementation of DApps in various domains have been investigated. Pierro et al. [21] discussed the adoption of Solidity as the most widely used programming language for coding DApps on the Ethereum blockchain.

### C. Parsing Technologies

ANTLR4 has proven to be a versatile and widely used parsing technology in various domains. Its features, such as predicates and support for LL(k) grammars, make it a powerful tool for parsing and analyzing different types of input, including source code, natural language, and more. ANTLR has been utilised as a back-end for Solidity parsers [22] and software vulnerabilities detectors [1], [17], [25].

Paso [22] is a web-based solidity parser that collects widely used software metrics from Solidity contracts. It relies on the same Solidity grammar, as MindTheDApps. Other tools use ANTLR to generate XML-based intermediate representation of smart contracts written in Solidity. For instance, SESCon (Secure Ethereum Smart Contracts by Vulnerable Patterns' Detection) [1] uses static analysis by converting a .sol file to its equivalent AST XML parse tree and apply the XPath query to find some simple vulnerabilities patterns. By combining XPath and taint analyses, SESCon can identify security vulnerabilities defined by the Ethereum community. Similarly, [17], [25] uses the Solidity parser to transform the smart contract source code into an XML parse tree which is then analysed further using XPath queries on the intermediate representation.

### D. Complex Networks

The use of complex networks to understand system behavior is well-established in various contexts, especially in object-oriented software systems. Gao et al. [10] employ directed software coupling networks to empirically analyze the macroscopic properties of such systems. Complex network analysis has been valuable not only in understanding software structure but also in real-time distributed control applications, where fast processes and complex interactions are key.

The method is also effective for assessing software risk and vulnerabilities. Cai et al. [6] use complex network analysis for vulnerability detection method based on deep learning and subgraph partition that enhances detection accuracy while maintaining scalability. Additionally, complex network models shed light on the behavior and emergence of requirements in networked systems [13], and can even guide software design and performance improvement.

Ferretti et al. [9] conducted an analysis of the Ethereum blockchain using complex network modeling techniques. They represented the flow of transactions in the blockchain as a network, with nodes representing Ethereum accounts. This approach allowed them to gain insights into the structure and dynamics of the Ethereum blockchain.

These studies highlight the versatility of complex network analysis in gaining insights into the structure, behavior, and vulnerabilities of software systems. Such insights are particularly relevant for decentralised applications, where understanding the interactions among smart contracts is essential.

## X. FUTURE WORKS

Future work will focus on broadening the dataset of analyzed DApps to deepen our understanding of smart contract interactions within decentralized applications. DAppRadar offers valuable data, including new releases and trending applications across categories, which could be useful for more comprehensive studies. Currently, our research focuses on Ethereum-based DApps, giving us a specific view of smart contract interactions. To provide a wider picture, we plan to extend our analysis to DApps built on other platforms like EOS, Solana, Hyperledger, and Cardano. Each of these platforms uses its own programming languages for smart contract development, necessitating the creation of new parsers for each. For example, while Ethereum primarily uses Solidity, Hyperledger employs languages like JavaScript and Python.

In addition, we aim to study the evolution of DApps within the Ethereum ecosystem by comparing older and newer applications. Such a comparative analysis would allow us to understand changes in DApp structures and smart contract interactions over time.

## XI. CONCLUSIONS

In this paper, we have introduced a tool designed to highlight interactions among smart contracts in decentralized applications. The tool accomplishes this by parsing the Abstract Syntax Tree to extract various elements including contract calls, modifiers, constructors, and events. This extraction offers a more complete understanding of the decentralized application being analyzed. The tool serves as a valuable resource for both developers and researchers aiming to grasp the purpose and structure of a decentralised application and the interactions among its smart contracts, and the interactions with external libraries. By providing these insights, our tool opens the door for more in-depth analysis and improvement of smart contracts and DApps development, contributing to the evolving landscape of blockchain technology.